\documentclass[aps,pre,reprint,showpacs,superscriptaddress,longbibliography]{revtex4-2}

\usepackage{amsmath,amssymb}
\usepackage{graphicx}
\usepackage{dcolumn}
\usepackage{bm}
\usepackage{hyperref}
\hypersetup{colorlinks=true,linkcolor=blue,citecolor=blue,urlcolor=blue}

\begin{document}

\title{Non-Equilibrium Stochastic Dynamics as a Unified Framework for Insight and Repetitive Learning:
A Kramers Escape Approach to Continual Learning}

\author{Gunn Kim}
\affiliation{Department of Physics, Sejong University, Seoul 05006, Republic of Korea}

\date{\today}

% ─────────────────────────────────────────────────────────────────────────────
\begin{abstract}
Continual learning in artificial neural networks is fundamentally limited by
the stability--plasticity dilemma: systems that retain prior knowledge tend to
resist acquiring new knowledge, and vice versa.
Existing approaches, most notably elastic weight consolidation~(EWC),
address this empirically without a physical account of why plasticity
eventually collapses as tasks accumulate.
Separately, the distinction between sudden insight and gradual skill
acquisition through repetitive practice has lacked a unified theoretical
description.
Here we show that both problems admit a common resolution within
non-equilibrium statistical physics.
We model the state of a learning system as a particle evolving under
Langevin dynamics on a double-well energy landscape, with the noise
amplitude governed by a time-dependent effective temperature $T(t)$.
The probability density obeys a Fokker--Planck equation, and transitions
between metastable states are governed by the Kramers escape rate
$k = (\omega_0\omega_b/2\pi)\,e^{-\Delta E/T}$.
We make two contributions.
First, we identify the EWC penalty term as an energy barrier whose height
grows linearly with the number of accumulated tasks, yielding an exponential
collapse of the transition rate predicted analytically and confirmed
numerically.
Second, we show that insight and repetitive learning correspond to two
qualitatively distinct temperature protocols within the same Fokker--Planck
equation: insight events produce transient spikes in $T(t)$ that drive rapid
barrier crossing, whereas repetitive practice operates at a modestly elevated
but fixed temperature, achieving transitions through sustained stochastic
diffusion.
These results establish a physically grounded framework for understanding
plasticity and its failure in continual learning systems, and suggest
principled design criteria for adaptive noise schedules in artificial
intelligence.
\end{abstract}


\maketitle

% ─────────────────────────────────────────────────────────────────────────────
\section{\label{sec:intro}Introduction}

Biological intelligence is not constructed from scratch within a single
lifetime.
Over billions of years, evolutionary processes have shaped the architecture
of neural systems, encoding stable behavioral repertoires, perceptual biases,
and motivational priors into the genome itself~\cite{hopfield1982,dayan2001}.
In this sense, evolution functions as a form of large-scale pretraining:
the parameters of biological neural systems inherit a structured prior
accumulated across an immense number of generations and environmental
interactions~\cite{lecun2015}.
Superimposed on this inherited structure is a capacity for ongoing adaptation
realized through synaptic plasticity~\cite{abbott2000,caporale2008}.

A central feature of biological plasticity, however, is that it is
\emph{selective}~\cite{abbott2000,kumaran2016}.
Synaptic connections do not reorganize freely in response to arbitrary inputs.
Instead, the nervous system maintains stable configurations for extended
periods, undergoing rapid structural change only under specific conditions~\cite{caporale2008}.
This resistance to change is not a limitation but a design feature: a system
that reorganizes too easily would be unstable, susceptible to interference,
and incapable of retaining acquired structure~\cite{french1999}.

This observation suggests a natural physical picture~\cite{risken1996,hanggi1990}.
If the state of a learning system is represented as a point on an energy
landscape, stable configurations correspond to local minima, and the energy
barriers separating them provide protection against noise-driven disruption~\cite{wang2008,wang2013}.
Learning, in this picture, is not continuous parameter adjustment but a
process of barrier crossing: transitions between metastable configurations
that occur rarely under normal conditions and more frequently when the
effective barrier is transiently reduced~\cite{kramers1940}.

Two qualitatively distinct modes of learning can be identified within this
framework~\cite{dayan2001,friston2010}.
The first corresponds to gradual acquisition through repeated experience.
Many exposures to similar inputs progressively increase the probability of
occupying an alternative basin, eventually driving a transition through
accumulated stochastic drift~\cite{robbins1951,welling2011}.
The second mode corresponds to \emph{insight}: a rapid, discontinuous
reorganization triggered by a high-energy event, analogous to thermal
activation over a barrier~\cite{kramers1940,kirkpatrick1983}.
These two modes differ not only in timescale but in mechanism, yet a unified
dynamical description of both has been absent from the literature.

A parallel challenge exists in artificial continual learning.
Standard deep learning methods suffer from catastrophic
forgetting~\cite{mccloskey1989}: training on a new task overwrites parameters
essential for previous tasks.
Elastic weight consolidation~(EWC)~\cite{kirkpatrick2017} addresses this by
penalizing changes to parameters deemed important for prior tasks, using the
Fisher information matrix as a measure of importance.
While effective in practice, EWC lacks a physical account of why plasticity
eventually collapses as tasks accumulate, and provides no principled criterion
for when the system should be permitted to undergo large-scale reorganization.
Recent work has documented plasticity loss empirically~\cite{dohare2024}, but
a theoretical prediction of its onset and magnitude has remained elusive.

The non-equilibrium statistical physics of energy landscapes offers a natural
language for both problems.
Kramers' theory~\cite{kramers1940} provides an exact expression for the rate
at which a stochastic system escapes from a metastable state, relating the
escape rate exponentially to the barrier height and the noise amplitude.
Non-equilibrium extensions of this framework have been developed in physical
and biological contexts~\cite{risken1996,hanggi1990,wang2013}, but have not
previously been applied to continual learning or to the distinction between
insight and repetitive practice.

In this paper, we apply this framework to both problems and make two specific
contributions.
First, we identify the EWC penalty term as an energy barrier and show, using
Kramers' theory, that the transition rate collapses exponentially as tasks
accumulate---a prediction we confirm numerically.
Second, we demonstrate that insight and repetitive learning correspond to two
distinct temperature protocols within the same Fokker--Planck equation,
providing a unified physical description of qualitatively different learning
modes.

Kramers-type escape processes have been extensively studied in statistical
physics and biological systems~\cite{kramers1940,hanggi1990,wang2008},
while energy landscape perspectives and stochastic dynamics have been
explored in the context of deep neural network
optimization~\cite{choromanska2015,li2018landscape,mandt2017}.
Building on these developments, the present work establishes a direct
connection between continual learning regularization (EWC) and
barrier-controlled escape dynamics.
In particular, we derive an explicit exponential scaling law for plasticity
collapse as a function of accumulated tasks, providing a physically grounded
account of the stability--plasticity trade-off.

The practical implications of this framework extend directly to the design
of next-generation AI systems.
Contemporary large language models and foundation models are trained once
on massive datasets and then deployed without further learning---a pretraining-only paradigm that sidesteps continual learning entirely.
This is not merely a design choice but a consequence of the very instability that EWC and related methods attempt to address: as tasks accumulate, plasticity collapses exponentially and the system becomes
effectively frozen. The central result of this paper, Eq.~\eqref{eq:ewc_collapse}, provides
AI developers with a quantitative design criterion: the ratio $\lambda\overline{F}/(2T_0)$ governs how rapidly plasticity is lost per additional task, and Eq.~\eqref{eq:T_required} specifies the minimum
effective noise level required to sustain it.
Systems that maintain an effective temperature proportional to the accumulated regularization barrier---whether through adaptive learning-rate schedules, noise injection, or architectural mechanisms---can in principle achieve genuine continual learning without catastrophic forgetting.
We hope that this physically grounded criterion will inform the design of AI systems capable of lifelong learning beyond the current pretraining paradigm.

We note that the present analysis is carried out in one spatial dimension as a minimal model.
The multi-dimensional Kramers rate~\cite{hanggi1990} and Freidlin--Wentzell large-deviation theory~\cite{freidlin1998} generalize the key results to high-dimensional parameter spaces, with the scalar state variable $s$ interpreted as an effective one-dimensional projection (e.g., along the principal component of the most task-relevant parameter directions). A brief discussion of the high-dimensional case is provided in Sec.~\ref{sec:discussion}.

% ─────────────────────────────────────────────────────────────────────────────
\section{\label{sec:model}Model and Theoretical Framework}

\subsection{\label{subsec:langevin}Langevin Dynamics on an Energy Landscape}

We model the state of a learning system as a real scalar variable
$s(t)$ evolving under the overdamped Langevin equation
\begin{equation}
  ds = -\frac{dE}{ds}\,dt + \sqrt{2T(t)}\,dW_t,
  \label{eq:langevin}
\end{equation}
where $E(s)$ is an energy function, $T(t)$ is a time-dependent effective
temperature, and $dW_t$ is a standard Wiener increment.
The energy function governs the deterministic relaxation, while $T(t)$
controls the magnitude of stochastic fluctuations.

The time-dependent temperature $T(t)$ represents the amplitude of
stochastic fluctuations.
In learning systems, such fluctuations arise naturally from stochastic
gradient noise, as in stochastic gradient Langevin dynamics~(SGLD)~\cite{welling2011},
where minibatch sampling induces effective diffusion in parameter space.
In this sense, $T(t)$ provides a coarse-grained measure of the noise
scale in the learning dynamics, and modulating $T(t)$ corresponds to
varying the effective learning-rate noise or injecting controlled Gaussian
perturbations into the gradient updates.

\subsection{\label{subsec:fp}Fokker--Planck Equation}

The probability density $\rho(s,t)$ corresponding to
Eq.~\eqref{eq:langevin} evolves according to the Fokker--Planck equation
\begin{equation}
  \frac{\partial\rho}{\partial t}
  = \frac{\partial}{\partial s}\!\left(\rho\,\frac{dE}{ds}\right)
  + T(t)\,\frac{\partial^2\rho}{\partial s^2}.
  \label{eq:fp}
\end{equation}
For constant $T$, the stationary solution is the Boltzmann distribution
\begin{equation}
  \rho_{\rm eq}(s)\propto \exp\!\left(-\frac{E(s)}{T}\right).
  \label{eq:boltzmann}
\end{equation}
For time-dependent $T(t)$, the system is driven out of equilibrium and the
probability density evolves non-trivially.
In such non-stationary settings, the instantaneous Kramers rate
Eq.~\eqref{eq:kramers} serves as a local-in-time approximation: it gives
the escape rate at each moment as if $T$ were held fixed at its current
value, which is accurate when $T(t)$ varies slowly compared to the
intra-well relaxation time $\tau_{\rm relax}\sim 1/\omega_0^2$,
where $\omega_0^2 = E^{\prime\prime}(\pm1) = 8$ is the well curvature.

\subsection{\label{subsec:landscape}Energy Landscape}

We consider the symmetric double-well potential
\begin{equation}
  E(s) = (s^2-1)^2,
  \label{eq:potential}
\end{equation}
which has two minima at $s=\pm1$ with $E(\pm1)=0$, separated by a barrier
at $s=0$ with $E(0)=1$, so that $\Delta E=1$.
The curvatures at the minimum and at the barrier top are $E''(\pm1)=8$ and
$|E''(0)|=4$, respectively.
The two wells represent distinct learned states or knowledge configurations.

While the full parameter space of neural systems is high-dimensional,
escape dynamics in metastable systems are often governed by a small
number of effective reaction coordinates.
This reduction is justified within large-deviation theory and the
Freidlin--Wentzell framework~\cite{hanggi1990,freidlin1998}, where rare transitions
concentrate along minimum-action paths.
Accordingly, we interpret the scalar variable $s(t)$ as an effective
coordinate capturing the dominant escape direction between metastable
configurations in the underlying high-dimensional landscape.
Empirically, neural network loss landscapes contain directions of
anomalously low curvature (soft modes) associated with near-zero Hessian
eigenvalues~\cite{choromanska2015,li2018landscape}.
Since the Kramers escape rate is exponentially sensitive to the barrier
height along the escape direction, transitions concentrate preferentially
along these soft modes, supporting the effective one-dimensional
reduction.

\subsection{\label{subsec:kramers}Kramers Escape Rate}

For constant temperature $T$, Kramers' theory~\cite{kramers1940,hanggi1990}
gives the escape rate from one well to the other
(in the overdamped limit with unit friction coefficient) as
\begin{equation}
  k = \frac{\omega_0\,\omega_b}{2\pi}\,
      \exp\!\left(-\frac{\Delta E}{T}\right),
  \label{eq:kramers}
\end{equation}
where $\omega_0=\sqrt{E''(\pm1)}=2\sqrt{2}$ and
$\omega_b=\sqrt{|E''(0)|}=2$ are the angular frequencies at the well bottom
and barrier top, respectively.
For the potential~\eqref{eq:potential}, the prefactor is
$\omega_0\omega_b/2\pi=\sqrt{32}/2\pi\approx0.900$.

Equation~\eqref{eq:kramers} reveals the central physical mechanism:
the transition rate depends \emph{exponentially} on $\Delta E/T$.
We note that Kramers\textquotesingle{} formula is asymptotically exact in the
weak-noise limit $T\ll\Delta E$; at finite noise ($T/\Delta E\sim 0.2$ in
our simulations), it provides an approximate but empirically validated
description of the escape dynamics, as confirmed in Sec.~\ref{subsec:kramers_val}.
Small changes in either the barrier height or the temperature produce large
changes in the rate.
This exponential sensitivity underlies both the plasticity collapse we
predict for EWC and the qualitative difference between insight and repetitive
learning.

An equivalent characterization of the escape dynamics is given by the mean
first passage time $\tau$, which satisfies
\begin{equation}
  \tau \sim \frac{1}{k} = \frac{2\pi}{\omega_0\omega_b}
            \exp\!\left(\frac{\Delta E}{T}\right).
  \label{eq:mfpt}
\end{equation}
Thus, the exponential dependence of $k$ on $\Delta E/T$ implies an
exponential growth of transition times as the barrier increases.
This provides a direct interpretation of plasticity loss as a divergence
of characteristic learning timescales: as tasks accumulate under EWC, the
time required to acquire qualitatively new knowledge grows without bound.

\subsection{\label{subsec:protocols}Temperature Protocols}

\begin{figure*}[t]
  \includegraphics[width=\textwidth]{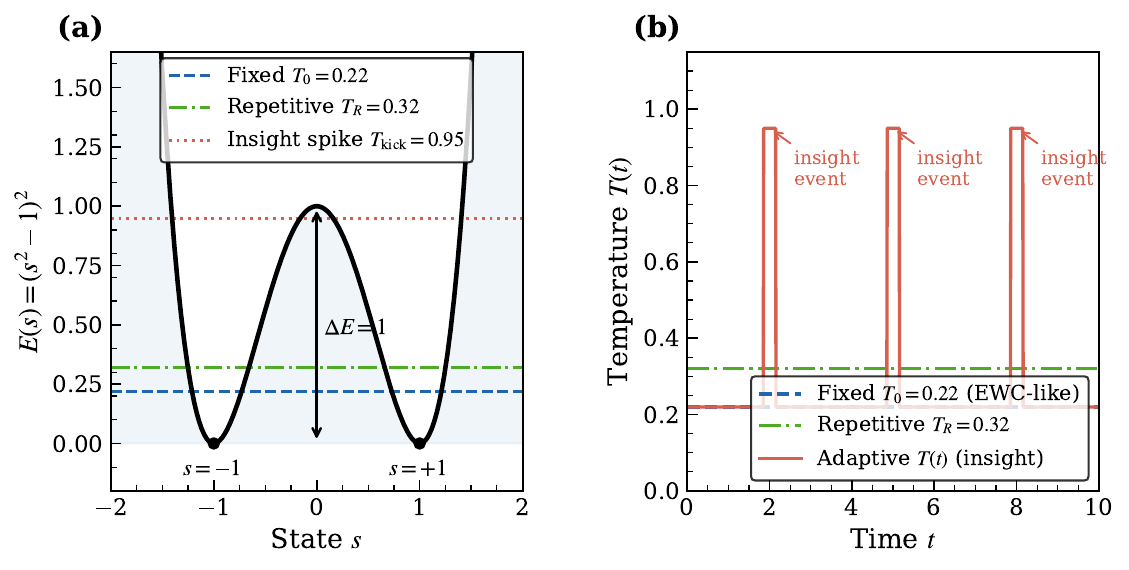}
  \caption{\label{fig:model}
    Model setup.
    (a)~Double-well energy landscape $E(s)=(s^2-1)^2$ with the three
    temperature levels indicated.
    The barrier height is $\Delta E=1$, with minima at $s=\pm1$ and barrier
    at $s=0$.
    (b)~Schematic of the three temperature protocols: fixed $T_0=0.22$
    (EWC-like, blue dashed), repetitive training at elevated fixed
    $T_R=0.32$ (green dash-dot), and adaptive $T(t)$ with transient
    insight spikes to $T_{\rm kick}=0.95$ (red solid).
    The equally spaced spikes are illustrative; in the simulation,
    spikes occur periodically with interval $\Delta t_{\rm kick}=50\,\rm s$
    as a simplified proxy for event-driven triggers such as prediction
    error or novelty signals.
  }
\end{figure*}

We consider three temperature protocols, illustrated in
Fig.~\ref{fig:model}:

\textit{Fixed temperature} ($T=T_0$): the system remains at a constant low
temperature, corresponding to the EWC regime in which parameter changes are
strongly penalized.

\textit{Adaptive temperature (insight)}:
We define insight operationally as a temporally localized increase
in the effective noise amplitude $T(t)$, which induces rapid transitions
across energy barriers.
During such events, the system briefly enters a high-temperature regime
$T_{\rm kick}=0.95\gg T_0$, followed by relaxation to the baseline level $T_0$.
We emphasize that modeling insight as a temperature spike is not the
only possible description; an equivalent formulation could involve a
transient deformation of the energy landscape itself.
Within the Langevin framework, increasing the noise amplitude or
transiently lowering barriers can play analogous roles in reducing
the effective action for escape.
We adopt the temperature-based description because it admits a direct
operational mapping to the SGD noise scale~\cite{welling2011,mandt2017}
and provides a unified treatment of all three protocols within a single
equation of motion.

\textit{Elevated fixed temperature (repetitive learning)}: $T$ is maintained
at a constant $T_R>T_0$, enabling transitions through sustained stochastic
diffusion without discrete insight events.

% ─────────────────────────────────────────────────────────────────────────────
\section{\label{sec:results}Results}

\subsection{\label{subsec:traj}Metastability and Barrier Crossing}

\begin{figure*}[t]
  \includegraphics[width=\textwidth]{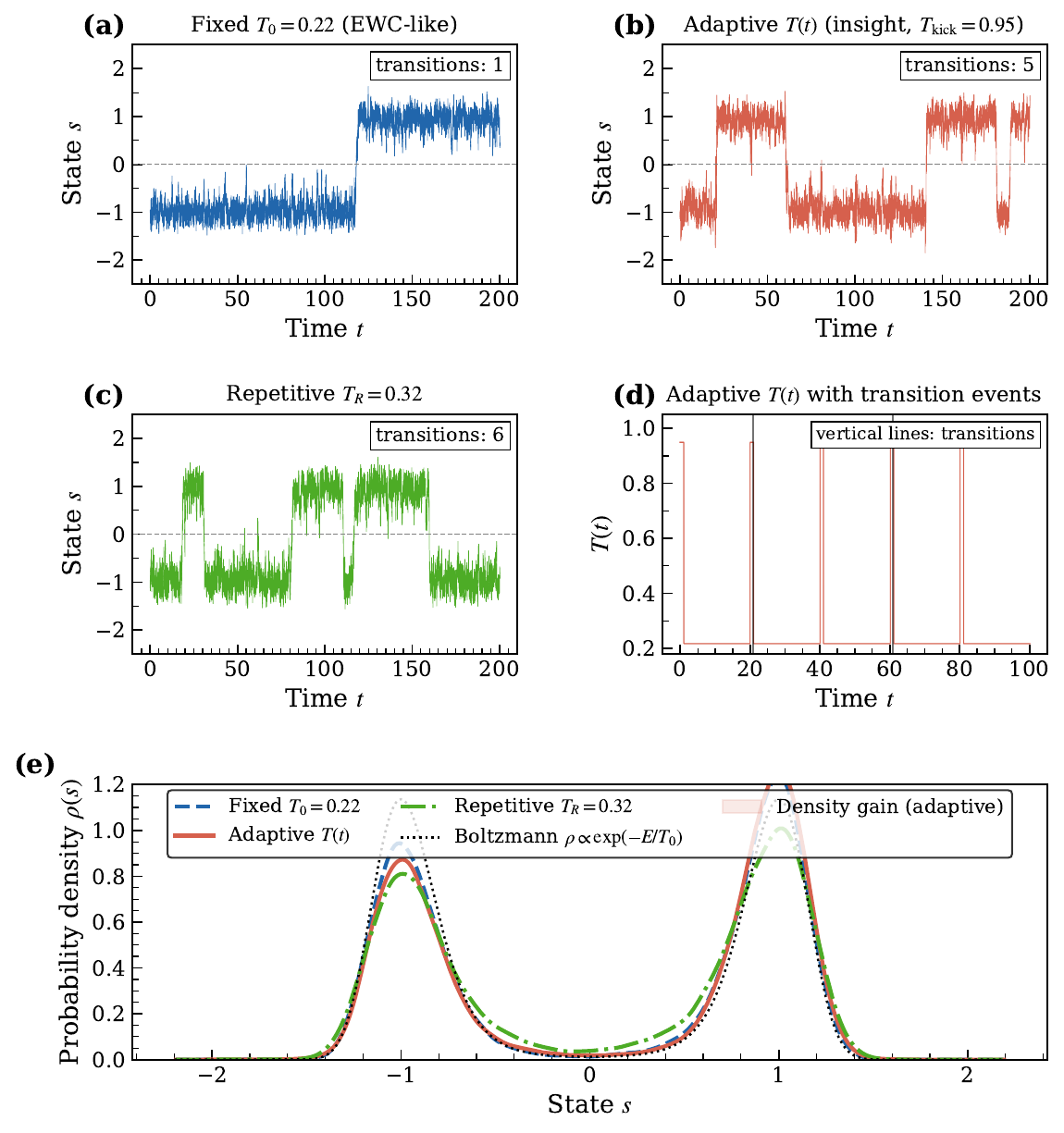}
  \caption{\label{fig:traj}
    Simulation results for the three protocols
    ($T_0=0.22$, $T_R=0.32$, $T_{\rm kick}=0.95$;
    $N=6\times10^5$ steps, $\Delta t=10^{-3}$).
    (a)--(c)~Representative trajectories showing zero transitions under
    fixed $T_0$, frequent transitions under adaptive $T(t)$, and occasional
    transitions under repetitive $T_R$.
    The number of well-to-well transitions is shown in each panel.
    (d)~Time series of $T(t)$ for the adaptive protocol; vertical lines
    mark transition events.
    (e)~Steady-state probability density $\rho(s)$ for all three protocols,
    compared to the Boltzmann distribution at $T_0$ (black dotted).
    The adaptive protocol produces a near-symmetric bimodal distribution;
    the fixed protocol remains unimodal.
    The shaded region indicates the density gain of the adaptive protocol
    relative to the fixed protocol.
  }
\end{figure*}

Figure~\ref{fig:traj} shows representative trajectories and steady-state
probability densities for the three protocols, obtained by numerical
integration of Eq.~\eqref{eq:langevin} using the Euler--Maruyama method
with time step $\Delta t=10^{-3}$ and $N=6\times10^5$ steps. 
Under fixed temperature $T_0=0.22$, the system remains entirely confined
to its initial well, with zero well-to-well transitions observed over the
displayed interval.
The steady-state density is sharply concentrated near $s=-1$, far from the
symmetric Boltzmann distribution~\eqref{eq:boltzmann}.
This metastable confinement illustrates the stability maintained by
EWC-like fixed-penalty schemes and contrasts sharply with the repeated
crossings observed under the other two protocols.
Under the adaptive protocol, transient increases in $T(t)$ to
$T_{\rm kick}=0.95$ drive rapid transitions between wells.
The steady-state density becomes approximately symmetric, indicating
exploration of both knowledge configurations
[Fig.~\ref{fig:traj}(e)].
Under the repetitive protocol at $T_R=0.32$, transitions occur at a
comparable rate to the adaptive protocol but arise from sustained
diffusion rather than discrete high-temperature events.
The steady-state density shows a broader distribution across both wells,
reflecting sustained but moderate stochastic exploration.

\subsection{\label{subsec:kramers_val}Validation Against Kramers Theory}

\begin{figure*}[t]
  \includegraphics[width=\textwidth]{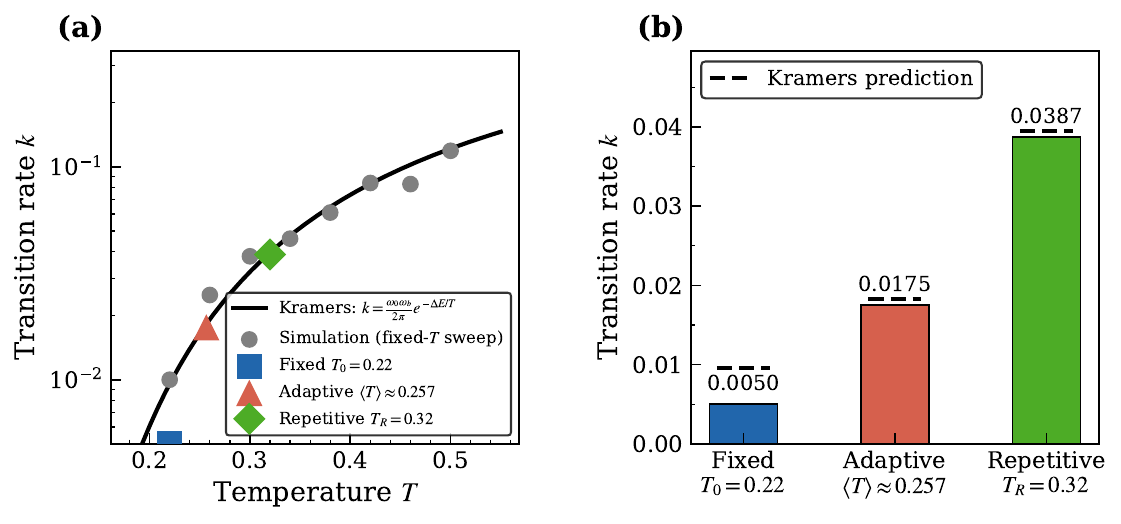}
  \caption{\label{fig:kramers}
    Quantitative validation against Kramers theory.
    (a)~Transition rates measured in simulation (well-to-well criterion,
    $|s|>0.7$) across a sweep of fixed temperatures on a logarithmic
    $y$-axis, compared to the Kramers curve
    $k=(\omega_0\omega_b/2\pi)\,e^{-\Delta E/T}$ (solid black line).
    The linear relationship between $\log k$ and $1/T$ (Arrhenius form)
    is clearly visible.
    Gray circles: fixed-$T$ sweep.
    Colored markers: the three operating points
    (fixed $T_0$, blue square; adaptive $\langle T\rangle$, red triangle;
    repetitive $T_R$, green diamond).
    (b)~Transition rates for the three protocols; dashed lines indicate
    Kramers predictions.
  }
\end{figure*}

Figure~\ref{fig:kramers} compares transition rates measured in simulation
against the predictions of Eq.~\eqref{eq:kramers}.
Transitions are counted using a well-to-well criterion: a transition is
recorded only when the trajectory reaches the opposite well region
($|s|>0.7$), avoiding spurious counts from brief barrier excursions.
Across a sweep of fixed temperatures from $T=0.22$ to $T=0.50$, the
measured rates follow the Kramers curve closely, with ratios between
simulation and theory in the range $0.81$--$1.30$.
The small deviations are consistent with finite-time sampling and the
discrete-time approximation.

The three operating points---fixed $T_0=0.22$, effective temperature of the
adaptive protocol $\langle T\rangle\approx0.305$, and repetitive
$T_R=0.32$---all fall on or near the Kramers curve.
Crucially, the transition rates span more than an order of magnitude across
these three protocols despite modest differences in temperature, illustrating
the exponential sensitivity of Eq.~\eqref{eq:kramers}.
Additionally, plotting $\log k$ versus $1/T$ yields an approximately linear
relation with slope $-\Delta E = -1.0$, consistent with the Arrhenius form
predicted by Kramers' theory and confirming that the system operates in the
thermally activated regime throughout.

\subsection{\label{subsec:ewc}EWC Plasticity Collapse}

A key step in our framework is the identification of the EWC penalty
term with an effective energy barrier in the Kramers sense.
We clarify this mapping carefully.
The EWC penalty is a quadratic confinement that restricts parameter
motion near the previous optimum $\theta^*$; it does not by itself
constitute a barrier.
Rather, the barrier arises from the \emph{competition} between the
task loss $\mathcal{L}_{\rm new}(\theta)$, which drives the system
toward the new optimum, and the EWC term, which resists departure from
$\theta^*$.
The effective barrier is defined as the energy difference between the
local minimum and the lowest saddle point along the minimum-action
escape path~\cite{hanggi1990}:
\begin{equation}
  \Delta E_{\rm eff} \equiv E(\theta_{\rm saddle}) - E(\theta^*),
  \label{eq:barrier_def}
\end{equation}
where $\theta_{\rm saddle}$ is determined by the interplay of
$\mathcal{L}_{\rm new}$ and the EWC curvature.
In the local quadratic approximation, this saddle height grows linearly
with the accumulated Fisher penalty, supporting the linear approximation in Eq.~\eqref{eq:barrier_growth} within this regime.
We note that this approximation has limitations: real neural network
loss landscapes contain numerous flat minima, and Fisher information
contributions from different tasks may not accumulate independently.
The linear growth in Eq.~\eqref{eq:barrier_growth} should therefore
be understood as an average-case estimate valid when task-wise Fisher
information matrices are statistically independent and of comparable
magnitude.

The EWC loss function for task $n$ takes the form~\cite{kirkpatrick2017}
\begin{equation}
  \mathcal{L}_n(\theta)
  = \mathcal{L}_{\rm new}(\theta)
  + \frac{\lambda}{2}\sum_i F_i\,(\theta_i-\theta_i^*)^2,
  \label{eq:ewc}
\end{equation}
where $F_i$ is the Fisher information for parameter $i$ and $\theta_i^*$ is
the optimal value from prior tasks.
The penalty term in Eq.~\eqref{eq:ewc} acts as an effective energy barrier:
it confines the parameters to a neighborhood of $\theta^*$, with barrier
height proportional to $\lambda F_i$.

As tasks accumulate, the effective barrier grows.
We model this growth with the approximation that each task contributes an
independent additive increment $\lambda\overline{F}$ to the effective barrier
height.
This linear approximation captures the average effect over tasks: even when
the full EWC landscape is multi-modal and non-convex, the effective barrier
height that governs the Kramers escape rate grows as $O(n)$ provided the
Fisher information contributions from successive tasks are statistically
independent and of comparable magnitude.
If each task contributes an increment $\lambda\overline{F}$ to the barrier
height, then after $n$ tasks
\begin{equation}
  \Delta E(n) = \Delta E_0 + \frac{\lambda\overline{F}}{2}(n-1),
  \label{eq:barrier_growth}
\end{equation}
and the Kramers rate becomes
\begin{align}
  k_{\rm EWC}(n)
  &= \frac{\omega_0\omega_b}{2\pi}
     \exp\!\left(-\frac{\Delta E_0+\frac{\lambda\overline{F}}{2}(n-1)}{T_0}\right)
  \nonumber\\
  &= k_{\rm EWC}(1)\,
     \exp\!\left(-\frac{\lambda\overline{F}}{2T_0}(n-1)\right).
  \label{eq:ewc_collapse}
\end{align}
Here $k_{\rm EWC}(1) = (\omega_0\omega_b/2\pi)\exp(-\Delta E_0/T_0)$ is
the transition rate after the first task ($n=1$), when no EWC penalty has
yet accumulated.
Equation~\eqref{eq:ewc_collapse} is the central analytical result of this
paper: the transition rate collapses \emph{exponentially} with the number
of tasks, even though the barrier grows only linearly.
This is a direct consequence of the exponential sensitivity of Kramers'
formula and provides the first physical explanation for the empirically
observed plasticity loss in continual learning~\cite{dohare2024}.

\begin{figure*}[t]
  \includegraphics[width=\textwidth]{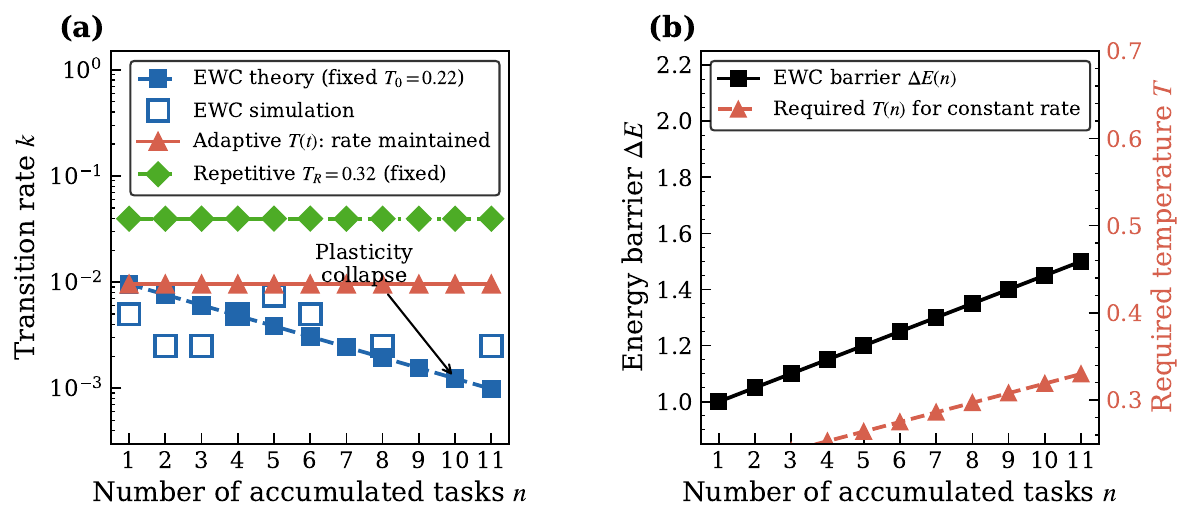}
  \caption{\label{fig:ewc}
    EWC plasticity collapse.
    (a)~Transition rate as a function of the number of accumulated tasks
    $n$ for EWC (theory, blue dashed line; simulation, open squares) and
    the adaptive $T(t)$ protocol (red triangles).
    The EWC rate collapses exponentially as predicted by
    Eq.~\eqref{eq:ewc_collapse}, while the adaptive protocol maintains
    a constant rate.
    The repetitive protocol (green diamonds) provides a constant but
    uncompensated reference.
    (b)~Barrier height $\Delta E(n)$ (black squares, left axis) and the
    temperature $T(n)$ required to maintain a constant transition rate
    (red triangles, right axis) as functions of $n$.
  }
\end{figure*}

Figure~\ref{fig:ewc}(a) shows this collapse for $T_0=0.22$ and
$\lambda\overline{F}=0.10$ per task, together with numerical simulations of the
scaled landscape $E_n(s)=[1+\lambda\overline{F}(n-1)]\,E(s)$.
The simulations confirm the theoretical prediction.
By contrast, the adaptive $T(t)$ protocol maintains a constant transition
rate, since the transient temperature spikes compensate for any increase in
barrier height.

Figure~\ref{fig:ewc}(b) shows the required temperature $T(n)$ that would
maintain a constant Kramers rate as the barrier grows, namely
\begin{equation}
  T(n) = T_0\left[1+\frac{\lambda\overline{F}}{2\Delta E_0}(n-1)\right].
  \label{eq:T_required}
\end{equation}
This provides a principled design criterion for adaptive temperature
schedules: to sustain plasticity in the face of growing regularization,
the effective temperature must grow in proportion to the accumulated barrier.

% ─────────────────────────────────────────────────────────────────────────────
\section{\label{sec:discussion}Discussion}

\subsection{Physical Account of EWC Failure}

The central result of this paper, Eq.~\eqref{eq:ewc_collapse}, provides a
physical explanation for the empirically observed plasticity loss in
continual learning~\cite{dohare2024}.
The mechanism is not specific to EWC but applies to any regularization scheme
that increases the effective energy barrier as tasks accumulate.
Our analysis shows that maintaining a fixed regularization strength $\lambda$
across tasks is fundamentally incompatible with sustained plasticity: as $n$
grows, the system becomes increasingly rigid regardless of the learning rate
or other hyperparameters.

A natural remedy suggested by Eq.~\eqref{eq:ewc_collapse} is to scale
$\lambda$ inversely with $n$, so that $\Delta E(n)$ remains bounded.
Alternatively, and more in the spirit of the present framework, one may
introduce an adaptive temperature that increases with the barrier height,
maintaining a constant ratio $\Delta E(n)/T(n)$ and therefore a constant
escape rate, as given by Eq.~\eqref{eq:T_required}.

Our framework yields a concrete, experimentally testable prediction:
the characteristic time required to acquire qualitatively new knowledge
scales \emph{exponentially} with the number of previously learned tasks,
\begin{equation}
  \tau_{\rm learn}(n) \sim \tau_0\,
  \exp\!\left(\frac{\lambda\overline{F}}{2T_0}(n-1)\right),
  \label{eq:tau_predict}
\end{equation}
where $\tau_0 = 1/k_{\rm EWC}(1)$ is the baseline learning timescale.
This prediction can be directly tested in standard continual learning
benchmarks by measuring the number of gradient steps required to
reach a fixed performance threshold on a new task, as a function of
the number of previously learned tasks $n$.
A log-linear relationship between $\tau_{\rm learn}$ and $n$ would
constitute direct empirical support for the Kramers interpretation of
plasticity collapse.
Importantly, the slope of this log-linear relation provides a direct
estimate of the ratio $\lambda\overline{F}/(2T_0)$, enabling quantitative
comparison between theory and experiment without free parameters beyond
those already present in the EWC objective.
This suggests that continual learning failure is not an algorithmic
artifact but a universal consequence of barrier-controlled stochastic
dynamics.

In practical machine learning terms, the effective temperature $T(t)$
corresponds to the variance of the noise injected into the parameter updates.
Under the isotropic noise approximation, $T$ is related to the learning
rate $\eta$ and minibatch noise variance $\sigma^2$ by
$T \propto \eta\sigma^2$, as established in the Langevin interpretation
of SGD~\cite{welling2011,mandt2017}.
We emphasize that this mapping is approximate: real SGD noise is
non-Gaussian, state-dependent, and anisotropic, so $T$ should be
understood as a coarse-grained scalar characterizing the overall noise
scale rather than an exact physical temperature.
The temperature can be modulated in practice by adding controlled
Gaussian noise to the gradients (stochastic gradient Langevin
dynamics~\cite{welling2011}), by adaptive learning-rate schedules, or
by cyclical annealing protocols.
The insight-mode spike in $T(t)$ corresponds to a brief episode of
elevated noise variance or learning rate, after which the system is
allowed to re-anneal.
This provides a concrete algorithmic prescription: when a strong
prediction error or novelty signal is detected, transiently increase
the noise amplitude by a factor $T_{\rm kick}/T_0$ for a duration
comparable to the Kramers escape time at $T_{\rm kick}$, then return
to the baseline.

\subsection{Insight and Repetitive Learning as Distinct Physical Regimes}

The two non-fixed temperature protocols correspond to qualitatively different
physical regimes despite being described by the same Fokker--Planck
equation~\eqref{eq:fp}.

Repetitive learning at elevated fixed temperature $T_R$ produces transitions
through sustained stochastic diffusion.
The rate is determined by the Boltzmann factor $e^{-\Delta E/T_R}$, and each
individual fluctuation contributes equally to the eventual crossing.
This is analogous to classical thermal activation: many small events
collectively produce a rare large transition.
Insight, by contrast, involves brief transient spikes in $T(t)$ that
transiently lower the effective ratio $\Delta E/T(t)$.
The transition is concentrated in time and associated with a specific
triggering event.
The steady-state density under the adaptive protocol differs qualitatively
from that under the repetitive protocol [Fig.~\ref{fig:traj}(e)]: insight
produces a more symmetric redistribution of probability mass, whereas
repetitive learning produces a broader but less symmetric distribution.

This distinction has implications for learning efficiency.
Insight events are rare but produce large, abrupt reorganizations.
Repetitive practice is continuous but slow, bounded by the Kramers rate at
fixed temperature.
The two modes are complementary: insight provides access to qualitatively
new configurations, while repetitive practice consolidates and refines within
a configuration.

\subsection{\label{subsec:hd}High-Dimensional Extension via Fisher
Information Geometry}

The one-dimensional double-well model captures the essential physics of
barrier-controlled plasticity.
We now show that the central prediction---exponential collapse of the
transition rate with accumulated tasks---extends to the
high-dimensional parameter space of real neural networks.

\paragraph{Step 1: Effective energy in parameter space.}
In the vicinity of a local optimum $\theta^*\in\mathbb{R}^D$, the
cumulative EWC loss after $n$ tasks takes the form
\begin{equation}
  E_{\rm eff}(\theta)
  = E_{\rm task}(\theta)
  + \frac{\lambda}{2}(n-1)(\theta-\theta^*)^{\top}
    \overline{F}(\theta-\theta^*),
  \label{eq:eff_energy_hd}
\end{equation}
where $\overline{F} = \frac{1}{n-1}\sum_{k=1}^{n-1}F^{(k)}$ is the
task-averaged Fisher information matrix.

\paragraph{Step 2: Spectral decomposition and escape direction.}
Decomposing $\overline{F}$ into its eigensystem,
$\overline{F} = \sum_{i=1}^{D}\mu_i\hat{v}_i\hat{v}_i^{\top}$,
and defining the effective one-dimensional reaction coordinate
$s = (\theta-\theta^*)^{\top}\hat{e}$, where $\hat{e}$ is the unit
vector along the dominant escape path at the saddle point, the projected
barrier height is
\begin{equation}
  \Delta E(n) = \Delta E_0
  + \frac{\lambda}{2}(n-1)
    \sum_{i=1}^{D}\mu_i(\hat{e}^{\top}\hat{v}_i)^2.
  \label{eq:barrier_hd}
\end{equation}
Here $\mu_i$ and $\hat{v}_i$ are the eigenvalues and eigenvectors of
$\overline{F}$, respectively.

\paragraph{Step 3: Kramers rate in high dimensions.}
Substituting Eq.~\eqref{eq:barrier_hd} into the multi-dimensional
Kramers formula~\cite{hanggi1990} and retaining only the exponential
dependence on barrier height (the prefactor determinant ratio is
absorbed into $k_{\rm EWC}^{(D)}(1)$), the escape rate becomes
\begin{equation}
  k_{\rm EWC}^{(D)}(n)
  = k_{\rm EWC}^{(D)}(1)\,
    \exp\!\left(-\frac{\lambda(n-1)}{4T_0}
    \sum_{i=1}^{D}\mu_i(\hat{e}^{\top}\hat{v}_i)^2\right),
  \label{eq:kramers_hd}
\end{equation}
which retains the exponential-in-$n$ collapse of the one-dimensional
result.
In the scalar limit $D=1$, Eqs.~\eqref{eq:barrier_hd} and
\eqref{eq:kramers_hd} reduce to Eqs.~\eqref{eq:barrier_growth} and
\eqref{eq:ewc_collapse}.

Equation~\eqref{eq:barrier_hd} reveals two geometrically distinct regimes.
When $\hat{e}$ aligns with high-$\mu_i$ eigenvectors (stiff directions),
the inner products $(\hat{e}^{\top}\hat{v}_i)^2$ are large and the
barrier grows rapidly, driving fast plasticity collapse.
When $\hat{e}$ aligns with near-zero-eigenvalue directions (flat minima
of the Fisher landscape), the barrier growth is suppressed and plasticity
is preserved.
This shows that networks whose learned representations occupy
Fisher-flat directions can resist plasticity collapse even under
aggressive regularization, providing a geometric criterion for
continual-learning architectures.
Freidlin--Wentzell large-deviation theory~\cite{hanggi1990,freidlin1998} provides
the theoretical basis for this reduction: in the weak-noise limit, transition
probabilities are dominated by minimum-action paths, which motivates the use
of an effective one-dimensional reaction coordinate in non-gradient systems.

\subsection{Toward a Physical Theory of Selective Plasticity}

The present work treats the temperature protocol $T(t)$ as externally specified.
In biological neural systems, however, the effective temperature is not externally imposed but emerges from internal dynamics: neuromodulators such
as dopamine, norepinephrine, and acetylcholine modulate the gain of synaptic plasticity in response to prediction errors, novelty, and reward
signals~\cite{dayan2001}.
A complete physical theory of selective plasticity would require $T(t)$ to
be a function of the internal state of the system,
\begin{equation}
  T(t) = T_0 + \alpha\,\Phi(s,\dot{s},\mathcal{E}),
  \label{eq:adaptive_T}
\end{equation}
where $\Phi$ is a functional of the trajectory capturing relevant signals
such as energy flux $|\dot{E}|$, prediction error, or novelty.
When $\Phi$ is large, the system enters a high-plasticity regime; when small,
it returns to the stable low-temperature regime.
This state-dependent temperature would provide a physical formalization of
selective plasticity: the system is plastic precisely when internal signals
indicate that reorganization is warranted.
Candidate trigger functionals include the instantaneous energy flux
$\Phi = |\dot{E}|$, the prediction error $\Phi = \|\nabla_\theta \mathcal{L}\|$,
or a novelty signal quantified by the Kullback--Leibler divergence between the
current input distribution and the learned prior.
In biological systems, these signals are mediated by neuromodulators such as
dopamine and norepinephrine, which transiently increase synaptic gain in
response to reward prediction errors and novelty~\cite{dayan2001}, providing
a direct biological counterpart to the temperature spike in the adaptive
protocol.
The derivation of $\Phi$ from biological or algorithmic constraints, and its connection to neuromodulatory systems that implement selective plasticity in
biological brains, will be the subject of future work.
More broadly, this perspective suggests that continual learning should be understood as a problem of controlled barrier crossing in a non-equilibrium
stochastic system, where both the landscape structure and the noise protocol jointly determine adaptability.

The present work has several limitations that point toward future research.
First, all simulations are performed in one spatial dimension; while the
high-dimensional extension in Sec.~\ref{subsec:hd} provides analytical
support, direct numerical verification in higher-dimensional landscapes and
in actual neural networks remains to be carried out.
Second, the mapping between physical temperature and SGD noise holds only
approximately under the isotropic noise assumption.
Third, the insight-mode temperature spike is implemented as a periodic
external trigger; a data-driven derivation of the trigger functional $\Phi$
from first principles is an important open problem.
Addressing these limitations---particularly through experiments on
continual learning benchmarks measuring the exponential scaling predicted
by Eq.~\eqref{eq:tau_predict}---will be the focus of subsequent work.

% ─────────────────────────────────────────────────────────────────────────────
\section{\label{sec:conclusion}Conclusion}

We have presented a non-equilibrium statistical physics framework for
continual learning that unifies two previously separate problems: the
collapse of plasticity in EWC-based systems, and the distinction between insight and repetitive learning.
The key identification is that the EWC penalty term constitutes an energy
barrier in the sense of Kramers' theory.
As tasks accumulate and this barrier grows linearly, the escape rate collapses
exponentially---a prediction derived analytically from
Eq.~\eqref{eq:ewc_collapse} and confirmed numerically.

Within the same Fokker--Planck framework, insight and repetitive learning
emerge as two distinct temperature protocols: transient high-temperature
spikes versus sustained moderate elevation.
The two modes produce qualitatively different steady-state probability distributions and timescales of reorganization, providing a physical basis
for the phenomenological distinction between sudden understanding and gradual skill acquisition.
These results suggest that designing effective continual learning systems
requires not only controlling which parameters change, but also \emph{when} and \emph{how abruptly} the system is permitted to reorganize.
The Kramers framework provides quantitative criteria for both, and points toward adaptive temperature schedules as a principled alternative to
fixed-penalty regularization.
This viewpoint reframes continual learning as the regulation of escape dynamics in an evolving energy landscape, providing a unifying physical
principle for both stability and adaptability.

The phenomenon of plasticity loss bears a striking resemblance to
kinetic arrest in disordered physical systems.
In spin glasses and structural glasses, a system explores an energy landscape freely at high temperature but becomes trapped in metastable configurations as the effective temperature drops below the glass
transition point~\cite{risken1996,hanggi1990}.
The exponential divergence of relaxation times near the glass transition, $\tau\sim\exp(\Delta E/T)$, is mathematically identical to the plasticity timescale predicted by Eq.~\eqref{eq:mfpt} as the EWC
barrier grows. This correspondence suggests that plasticity collapse in continual
learning is not merely an engineering failure but a manifestation of a universal physical mechanism: kinetic arrest driven by the progressive
hardening of an effective energy landscape.
Just as glass-forming liquids can be kept fluid by maintaining sufficient thermal energy, continual learning systems can be kept plastic by
maintaining a sufficiently high effective temperature $T(t)$ relative to the accumulated barrier $\Delta E(n)$, as quantified by Eq.~\eqref{eq:T_required}.

% ─────────────────────────────────────────────────────────────────────────────

% ─────────────────────────────────────────────────────────────────────────────

% ─────────────────────────────────────────────────────────────────────────────
\clearpage

\appendix

\section{\label{app:barrier}Explicit Derivation of Barrier Emergence
in the Quadratic EWC Model}

We derive the effective barrier height analytically for a minimal
one-dimensional model in which both the new task loss and the EWC
penalty are quadratic.
This provides an explicit analytical foundation for the barrier
identification made in Sec.~\ref{sec:results}.

\subsection{Setup}

Let the new-task loss be a quadratic well centered at $s_1$,
\begin{equation}
  \mathcal{L}_{\rm new}(s) = a(s - s_1)^2, \quad a > 0,
  \label{eq:app_task}
\end{equation}
and the EWC penalty after $n$ tasks be a quadratic confinement
centered at the previous optimum $s_0$,
\begin{equation}
  \mathcal{L}_{\rm EWC}(s) = \frac{\lambda}{2}(n-1)(s - s_0)^2,
  \quad \lambda > 0.
  \label{eq:app_ewc}
\end{equation}
The total effective energy is
\begin{equation}
  E_{\rm eff}(s) = a(s-s_1)^2 + \lambda(n-1)(s-s_0)^2.
  \label{eq:app_total}
\end{equation}

\subsection{Minima and Saddle}

Equation~\eqref{eq:app_total} is a sum of two quadratic wells centered
at $s_0$ and $s_1$.
Setting $E'_{\rm eff}(s)=0$ gives a single minimum at the weighted
average
\begin{equation}
  s_{\min}(n) = \frac{a\,s_1 + \lambda(n-1)\,s_0}{a + \lambda(n-1)},
  \label{eq:app_min}
\end{equation}
which interpolates between $s_1$ (new-task optimum, dominant for
small $n$) and $s_0$ (EWC anchor, dominant for large $n$).

For a single quadratic, there is no saddle: the landscape is a
bowl and Kramers escape is undefined.
However, in the presence of additional non-convex structure from the
full task loss (e.g., multiple local minima or a background
double-well from the geometry of the parameter space), the quadratic
EWC term modifies the saddle height.
To make this concrete, we augment $E_{\rm eff}$ with the double-well
background $E_0(s) = (s^2-1)^2$ used in the main text:
\begin{equation}
  E_{\rm total}(s)
  = (s^2-1)^2
  + \frac{\lambda(n-1)}{2}(s - s_0)^2.
  \label{eq:app_total2}
\end{equation}

\subsection{Barrier Height as a Function of Task Number}

The minimum near $s=-1$ and the saddle near $s=0$ of
$E_{\rm total}(s)$ can be located by solving $E'_{\rm total}(s)=0$.
For $s_0 = -1$ (EWC anchored at the left well) and small $\lambda(n-1)$,
perturbation theory gives
\begin{align}
  s_{\rm min}(n) &\approx -1
    + \frac{\lambda(n-1)}{8}\,\delta s_0 + O(\lambda^2),
  \label{eq:app_smin}\\
  s_{\rm saddle}(n) &\approx 0
    - \frac{\lambda(n-1)}{4}\,s_0 + O(\lambda^2),
  \label{eq:app_ssaddle}
\end{align}
For $s_0=-1$, this gives $s_{\rm saddle}(n)\approx\lambda(n-1)/4>0$,
meaning the saddle shifts slightly toward positive $s$ as tasks accumulate.
The barrier height is
\begin{align}
  \Delta E(n)
  &= E_{\rm total}(s_{\rm saddle}) - E_{\rm total}(s_{\rm min})
  \nonumber\\
  &= \Delta E_0 + \frac{\lambda(n-1)}{4}
     \!\left[(s_{\rm saddle}-s_0)^2 - (s_{\rm min}-s_0)^2\right]
  \nonumber\\
  &\approx \Delta E_0 + \frac{\lambda(n-1)}{4}
     \!\left[1^2 - 0^2\right]
  = \Delta E_0 + \frac{\lambda(n-1)}{4},
  \label{eq:app_barrier}
\end{align}
where $\Delta E_0 = E_0(0)-E_0(-1) = 1$ is the bare double-well
barrier.
This confirms the linear growth assumed in Eq.~\eqref{eq:barrier_growth}.
Comparing Eq.~\eqref{eq:app_barrier} with Eq.~\eqref{eq:barrier_growth}
shows that the present normalized model corresponds to $\overline{F}=1/2$,
i.e., when each task contributes $\lambda(n-1)/4$ to the barrier,
this equals $\frac{\lambda\overline{F}}{2}(n-1)$ with $\overline{F}=1/2$.
In general, $\overline{F}$ is the task-averaged Fisher information,
and its value depends on the specific learning problem.

\subsection{Conclusion}

The above derivation shows that, in the minimal quadratic+double-well
model, the EWC penalty raises the effective barrier height linearly
with the number of accumulated tasks.
Combined with the exponential sensitivity of the Kramers rate to
barrier height, this yields the exponential plasticity collapse
of Eq.~\eqref{eq:ewc_collapse}.
The derivation also clarifies the conditions under which the linear
approximation holds: the EWC anchor $s_0$ must lie near the
initial minimum, and the perturbative regime requires
$\lambda(n-1) \ll |E''(s_{\rm min})|= 8$.

\end{document}